\documentclass{iopart}
\usepackage{graphicx}
\usepackage{amsfonts}

\usepackage{psfrag}

\def\FigSize{6.45cm}
\def\be{\begin{equation}}
\def\ee{\end{equation}}
\def\ba{\begin{eqnarray}}
\def\ea{\end{eqnarray}}

\begin{document}
\def\be{\begin{equation}}
\def\ee{\end{equation}}
\def\ba{\begin{eqnarray}}
\def\ea{\end{eqnarray}}

\def\ket#1{|{#1}\rangle}

\title{DMRG study of the Berezinskii-Kosterlitz-Thouless
transitions of the 2D five-state clock model}

\author{Christophe Chatelain}
\address{
Groupe de Physique Statistique,
D\'epartement P2M,
Institut Jean Lamour (CNRS UMR 7198),
Universit\'e de Lorraine, France}

\ead{christophe.chatelain@univ-lorraine.fr}

\date{\today}

\begin{abstract}
The two Berezinskii-Kosterlitz-Thouless phase transitions of the
two-dimensional 5-state clock model are studied on infinite strips
using the DMRG algorithm. Because of the open boundary conditions, the
helicity modulus $\Upsilon_2$ is computed by imposing twisted magnetic
fields at the two boundaries. Its scaling behavior is in good
agreement with the existence of essential singularities
with $\sigma=1/2$ at the two transitions. The predicted universal
values of $\Upsilon_2$ are shown to be reached in the thermodynamic
limit. The fourth-order helicity modulus is observed to display
a dip at the high-temperature BKT transition, like the XY model,
and shown to take a new universal value at the low-temperature one.
Finally, the scaling behavior of magnetization at the
low-temperature transition is compatible with $\eta=1/4$.
\end{abstract}


\section{Introduction}
The critical behavior of the two-dimensional XY model has been the subject
of a huge literature. Because the symmetry group of the XY model is continuous,
the Mermim-Wagner-Hohenberg theorem states that long-range order is destroyed by
massless spin-wave excitations in dimension $d=2$~\cite{Mermin,Hohenberg}.
Therefore, in contrast to the Ising or 3 or 4-state Potts models,
there cannot exist any ferromagnetic phase, where the symmetry
would be spontaneously broken. However, at low temperature the two-dimensional
XY model presents a critical phase
that terminates at a topological Berezinskii-Kosterlitz-Thouless (BKT) phase
transition where the coherence of
spin waves is lost because of the proliferation of free topological
defects~\cite{Berezinski,Kosterlitz,Minnhagen}. BKT phase transitions were
found in a variety of context,
the superfluid transition of helium being certainly the most famous example of
such a transition.

However, there exist situations where the Mermim-Wagner-Hohenberg theorem can be
bypassed. For instance, the theorem extends to any kind of pair interaction
but not to hard-core potentials~\cite{Mermin68}. As a consequence, a two-dimensional
gaz of hard spheres crystallizes at low enough temperature. As the
temperature is increased, the system undergoes two successive phase transitions
separating, first the solid phase and an intermediate hexatic phase, and then,
the hexatic and liquid phases~\cite{Halperin}. The solid-hexatic phase transition
is a BKT transition while the nature of the hexatic-liquid transition is
controversial: it was long believed to be a BKT transition too but recent Monte Carlo
simulations provided evidence of a first-order phase transition~\cite{Krauth}.

Another way to escape from the Mermim-Wagner-Hohenberg theorem is to replace
the $U(1)$ symmetry of the XY model by a finite symmetry. This is the case of
the $q$-state clock model, originally introduced by Potts~\cite{Potts}.
The Hamiltonian is defined as
   \be H_0=-\sum_{(i,j)}\cos\Big({2\pi\over q}(\sigma_i-\sigma_j)\Big),\quad
   \sigma_i\in\{0,\ldots,q-1\}\label{H}\ee
where the sum extends over all pairs of nearest neighbors of the lattice. It is
invariant under the $\mathbb{Z}_q$ group of cyclic permutations of the $q$ states.
The XY model is recovered in the limit $q\rightarrow +\infty$. For $q\le 4$,
the two-dimensional clock model undergoes a unique second-order phase transition,
like the Potts model. For $q\ge 5$, the phase diagram shows, between the
ferromagnetic and paramagnetic phases, an intermediate critical phase with
quasi long-range order~\cite{Jose}. These three phases are separated by two BKT phase
transitions. The transition between the critical and paramagnetic phases occurs
at a temperature roughly independent of $q$.
The transition temperature of the second transition decays to zero as
the number of states $q$ goes to infinity, in agreement with the absence
of ferromagnetic order predicted by the Mermim-Wagner-Hohenberg theorem for the
2D XY model. The situation is however not completely clear in the case
$q=5$. Early Monte Carlo simulations provided evidence of the existence of three
distinct phases~\cite{Tobochnik}. The average complex magnetization was shown
to display the expected $\mathbb{Z}_5$-symmetry in the ferromagnetic
phase and a continuous $U(1)$ symmetry in the critical phase~\cite{Papa}.
If the low-temperature transition was recognized as a BKT transition,
the question of the nature of the high-temperature transition was subject
to controversy in the last years. The claim that the high-temperature
transition could not be of BKT-type for $q\le 6$~\cite{Lapilli} was
supported by a study of Fisher zeroes~\cite{Hwang}. However, recent Monte
Carlo studies showed that the helicity modulus $\Upsilon_2$, defined as the
response of the system to a twist~\cite{Fisher}, displays a jump, as
expected in the case of a BKT transition~\cite{Baek10,Papa2,Baek13,Kumano}.
The observation of a non-vanishing helicity modulus in the high-temperature
phase was explained as the result of an inappropriate definition of
$\Upsilon_2$ in the case of the clock model~\cite{Kumano}.

The helicity modulus is central in showing the BKT nature of the transition.
However, because it is defined as a derivative of the free energy with
respect to a twist, its computation is not straightforward in a Monte Carlo
simulation. In the references mentionned above, the boundary-flip Monte
Carlo method was employed~\cite{Gottlob}. In this paper, we present new
results for the second and fourth-order helicity modulus of the 5 and 6-state
clock models obtained using the Density Matrix Renormalization Group (DMRG)
algorithm~\cite{White1,White2,Schollwock}. Details about the numerical calculations
and the definition of the helicity modulus is presented in the second section.
The numerical data are analyzed in the third section. The scaling behaviour of
the helicity modulus, of the fourth-order helicity modulus and of magnetization
are compared to Kosterlitz-Thouless predictions. Conclusions follow.

\section{DMRG algorithm, observables and convergence}
In this section, the DMRG algorithm is very briefly presented. For more details
about the implementation of this algorithm, the interested reader is invited
to refer to the review~\cite{Schollwock}. Our implementation and the simulation
parameters are discussed. The definition of the estimator of the helicity modulus
is given and finally, the convergence is studied.

\begin{figure}[ht]
\centering
\includegraphics[width=10cm]{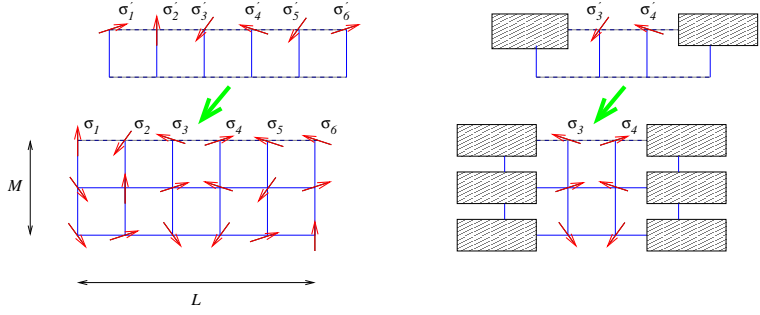}
\caption{On the left, schematic action of the transfer matrix on a $M\times L$
square lattice of the 5-state clock model. The spins on the upper row are denoted
as $\sigma_1$ to $\sigma_L$. A new row with spins $\sigma_1'$ to $\sigma_L'$
is added to the system, resulting into a $(M+1)\times L$ lattice. Denote
${\cal Z}_{M}(\sigma_1,\ldots,\sigma_L)$ the partition function of the
$M\times L$ lattice with fixed spins $\sigma_1$ to $\sigma_L$ on the upper row.
Similarly, ${\cal Z}_{M+1}(\sigma_1',\ldots,\sigma_L')$ is the partition
function after the addition of the $M+1$-th row. The linear relation
between them
   $${\cal Z}_{M+1}(\sigma_1',\ldots,\sigma_L')
   =\sum_{\sigma_1,\ldots,\sigma_L}
   T(\sigma_1',\ldots,\sigma_L';\sigma_1,\ldots,\sigma_L)
   {\cal Z}_{M}(\sigma_1,\ldots,\sigma_L)$$
suggests to use a matricial notation: $\ket{{\cal Z}_{M+1}}=T\ket{{\cal Z}_{M}}$
where $T$ is the transfer matrix acting in the $q^L$-dimensional vector space
of the spin configurations on the upper row of the two-dimensional system.
On the right, schematic representation of the action of the effective transfer matrix
$T_{\rm eff}$ constructed by the DMRG algorithm. Each vector of the vector space
on which acts $T_{\rm eff}$ corresponds to a configuration of the two central
spins and to effective degrees of freedom on both sides (represented as shaded
rectangles).}\label{tm}
\end{figure}

From the Hamiltonian (\ref{H}) on a square lattice, a transfer matrix $T$ is
easily defined in the vector space of spin configurations on the last row
of a strip of width $L$ (see figure~\ref{tm}). Because the hamiltonian couples
only nearest-neighbouring spins, the transfer matrix can be expressed as a
product of matrices, whose elements are the Boltzmann weights $T(\sigma',\sigma)
=e^{\beta J\cos {2\pi\over q}(\sigma'-\sigma)}$ of each new single bond betwen the
$\sigma$ and $\sigma'$. This property is important for DMRG.
The free energy of the clock-model on an infinite strip of width $L$
is proportional to the logarithm of the largest eigenvalue of the transfer
matrix $T$. However, since the spins take $q$
possible values, the dimension of this vector space grows as $q^L$, i.e.
exponentially fast. In the case $q=5$ we are interested in, this limits the
strip widths that can be studied to small values, $L=9$ or 10.
In the Density Matrix Renormalization Group (DMRG) approach, also called
Transfer Matrix Renormalization Group (TMRG) in the context of classical
systems, the transfer matrix $T$ is replaced by an effective transfer $T_{\rm eff}$
matrix acting in a much smaller vector space. The vectors of the latter
corresponds to the configurations of two spins surrounded by effective degrees
of freedom, the left and right blocks. $T_{\rm eff}$ can be constructed
iteratively, i.e. by considering strips of increasing width $L$.
For a small system, the eigenvector of the transfer matrix $T$ associated
to the largest eigenvalue can be computed exactly.
Effective transfer matrices with a smaller dimensions are then constructed
for the left and right blocks. As shown by White~\cite{White1,White2,
Schollwock}, the best approximation, in the sense of mean-square deviation
of the eigenvector, is to truncate the basis of the left and right blocks
by keeping only the eigenvectors associated to the largest eigenvalues
of the reduced density matrix of the block. The central spins are then
absorbed into the left and right blocks and two new spins are usually
introduced between them. The procedure is repeated until the desired strip
width is reached. This algorithm, called infinite-size DMRG, is efficient
for homogeneous system. In our case,
we will introduce magnetic fields at the boundaries. Therefore, we proceeded
differently: only one block, say left, is considered. The second block,
say right, is reduced to a single spin. After the construction of an
effective transfer matrix for the left block, new spins are not introduced
between the former central spins but at the right of the right block.
The procedure is iterated up to the desired strip width. The same algorithm
is used to construct a sequence of effective transfer matrices for the
right block. Once the final strip width $L$ has been reached, the accuracy of
the free energy can be further improved by applying several iterations of the
finite-size DMRG algorithm. A sweep over the system is performed by increasing
the length of one block and decreasing the length of the other one. We
systematically performed six sweeps, in both directions. Even for the
longest strips, the free energy was observed to reach a stable value
(10 digits were considered), after three sweeps.

Since the DMRG algorithm gives directly access to the free energy, the
helicity modulus $\Upsilon_2$ can be calculated as
   \be\Upsilon_2=L^2\left({\partial^2 f\over\partial\Delta^2}\right)_{\Delta=0}
   \label{DefUpsilon}\ee
where $f(\Delta)$ is the free energy density of the system when a twist
$\Delta$ is introduced at its boundary.
However, as emphasized by Kumano {\sl et al.}~\cite{Kumano}, spins are
not continuous in the clock model but takes a finite number of states 
so the twist can only be a multiple of $2\pi/q$. The usual definition
of the helicity modulus should be replaced in this case by
   \be\Upsilon_2={2L^2\over (2\pi/q)^2}\big[f(2\pi/q)-f(0)\big].
   \label{DefUpsilon2}\ee
The twist is usually introduced in the system by replacing the
interaction term in the Hamiltonian on one row by
   \be -\cos\Big({2\pi\over q}(\sigma_i-\sigma_j+1)\Big).\ee
With periodic boundary conditions, a spin wave is induced to accommodate
this interaction term. Unfortunately, periodic boundary conditions are
known to greatly deteriorate the convergence of the DMRG algorithm.
We therefore introduced a twist by imposing magnetic fields at the
two boundaries of an open system. The Hamiltonian now reads
   \be\fl H_\Delta=-\sum_{(i,j)}\cos\Big({2\pi\over q}(\sigma_i-\sigma_j)\Big)
   -\sum_{i\in S_L}\cos\Big({2\pi\over q}\sigma_i\Big)
   -\sum_{i\in S_R}\cos\Big({2\pi\over q}\sigma_i-\Delta\Big)\ee
where $S_L$ (resp. $S_R$) denotes the set of spins at the left (resp.
right) boundary of the strip. Equivalently, the system can be seen
as a strip of width $L+2$ with spins frozen in the states $\sigma_i=0$
and $\sigma_i={q\over 2\pi}\Delta$ on the first and last columns respectively.
The helicity modulus is determined as the finite difference of the free
energy~(\ref{DefUpsilon2}). We also tested the same definition but with a
larger twist $\Delta=4\pi/q$. Following~\cite{Kim03}, a fourth-order helicity
modulus $\Upsilon_4$ can be defined in the case of an infinitesimal twist
$\Delta$ from the Taylor expansion 
    \be f(\Delta)=f(0)+{\Delta^2\over 2}\Upsilon_2
    +{\Delta^4\over 4!}\Upsilon_4+{\cal O}(\Delta^6)\ee
of the free energy density. We introduced the estimator
   \be\Upsilon_4={2\over (2\pi/q)^4}\big[f(4\pi/q)+3f(0)-4f(2\pi/q)\big].
   \label{Y4}\ee
Finally, since the $\mathbb{Z}_q$ symmetry is broken, the spontaneous
magnetization can be computed when $\Delta=0$, i.e. when the magnetic
fields at the two boundaries are aligned in the same direction.

In Monte Carlo simulations, error bars on any observable $O$ are related
to the fluctuations of $O$. According to the central limit theorem, the error
on $O$ is given by $\sqrt{\sigma_O^2\tau/N}$ where $\sigma_O^2$ is the variance
of $O$, $N$ the number of Monte Carlo steps and $\tau$ the autocorrelation time.
Error bars are therefore purely statistical and a better accuracy is reached by
increasing the number of Monte Carlo steps. Systematic deviations arise
only when the system was not properly thermalized or when the dynamics is
not ergodic. In DMRG calculations, there is no statistical fluctuation of
free energy around its exact value but only systematic deviations
due to the truncation of the vector spaces of the left and right blocks.
These deviations are difficult to estimate.
In the following, the convergence of the helicity modulus is studied at
the inverse temperature $\beta=1/k_BT=1.08$, i.e. $T\simeq 0.925$. This
temperature roughly lies at the center of the critical phase of the
5-state clock model. It is therefore the most difficult case for
numerical simulations. Since the correlation length is expected to
diverge with the strip width $L$ in this phase, the left and right blocks
are strongly correlated through the two central spins. As a consequence,
the reduced density matrix is observed to have a slowly decaying
spectrum. Because the deviation of free energy is expected to be
proportional to the sum of the eigenvalues of the reduced density
matrix associated to the eigenvectors that have been
discarded~\cite{Legeza}, each truncation of the basis of the two blocks
leaves an effective transfer matrix that may reproduce poorly these
correlations if a too small number of states is kept. 

\begin{figure}[ht]
\centering
\psfrag{1/L}[Bc][Bc][1][1]{$1/m^4$}
\psfrag{Y}[Bc][Bc][1][1]{$\Upsilon_2$}
\psfrag{L=16}[Bc][Bc][1][0]{\tiny $L=16$}
\psfrag{L=24}[Bc][Bc][1][0]{\tiny $L=24$}
\psfrag{L=32}[Bc][Bc][1][0]{\tiny $L=32$}
\psfrag{L=64}[Bc][Bc][1][0]{\tiny $L=64$}
\psfrag{L=128}[Bc][Bc][1][0]{\tiny $L=128$}
\psfrag{L=256}[Bc][Bc][1][0]{\tiny $L=256$}
\psfrag{L=512}[Bc][Bc][1][0]{\tiny $L=512$}
\hbox{\includegraphics[width=\FigSize]{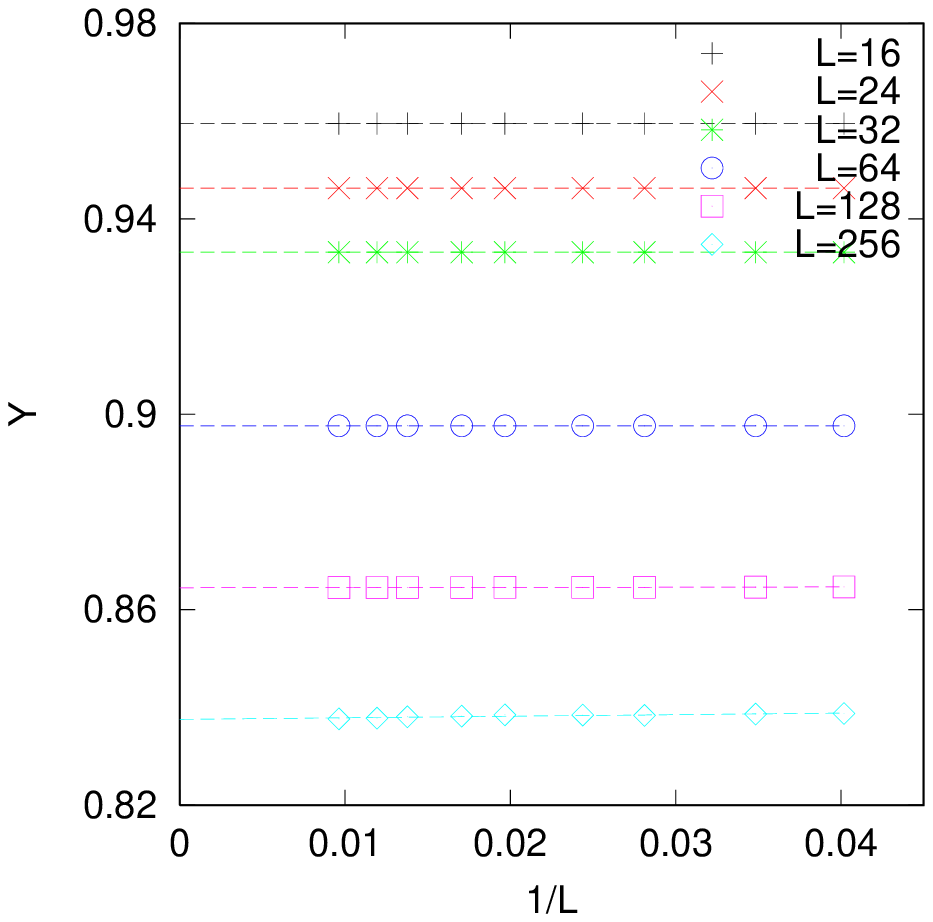}
\quad\vbox{\hsize=6.1cm
\includegraphics[width=6.1cm]{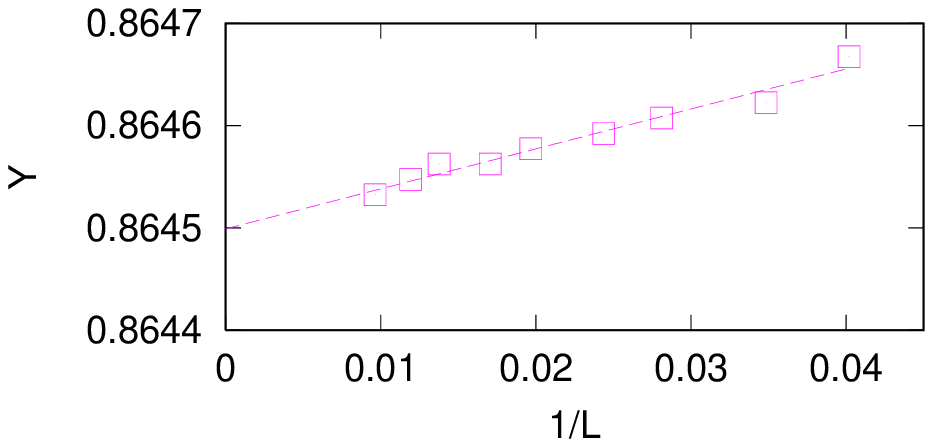}\par
\includegraphics[width=6.1cm]{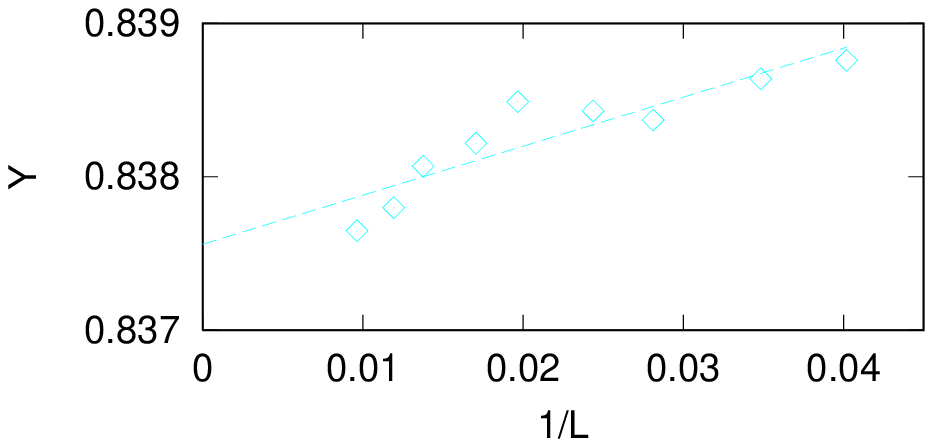}\par
}}
\caption{Helicity modulus $\Upsilon_2$ of the 5-state clock model at the
inverse temperature $\beta=1/k_BT=1.08$ versus $1/m^4$ where $m$ is
the number of states of the truncated vector spaces of the left and
right blocks during the DMRG process. The different curves correspond
to different strip widths $L$ and the straight lines to linear fits.
On the right, a zoom is performed on the curves for $L=128$ (top) and
$L=256$ (bottom).
}\label{Fig0}
\end{figure}

On figure~\ref{Fig0}, the helicity modulus is plotted versus the number
of states $m$ kept after truncation. More precisely, $m$ refers to the
number of states describing both the left (resp. right) block and the left
(resp. right) central spin after the renormalization step. The dimension
of the full vector space is therefore $m^2$. For the smallest strip widths $L$,
the numerical data do not show any dependence on $m$. However, the free
energy displays a small evolution at the 8th digit for $L\ge 64$. Since
the helicity modulus Eq.~\ref{DefUpsilon2} is defined as the difference of
two free energies, the deviation is larger for $\Upsilon_2$. Indeed, for
$L=64$, the free energies are of order ${\cal O}(10^4)$ while their difference
is of order ${\cal O}(10^0)$. As a consequence, an evolution of $\Upsilon_2$
with $m$ is observed at the 4th digit for $L=64$ and at the third one for
$L=256$. To quantify the systematic deviation of our data, we extrapolated
them to the limit $m\rightarrow +\infty$. As observed for instance in the case
of $SU(N)$ chains~\cite{Fuhringer}, the free energy is well described by a
power law decay $f(m)=f(\infty)+am^{-\alpha}$ with $\alpha\simeq 4$. The
helicity modulus $\Upsilon_2$ is also reasonnably well fitted by such a law,
though other laws, in particular $\Upsilon_2(m)=\Upsilon_2(\infty)+ae^{-m/m_0}$
with $m_0\simeq 70$, give sensibly similar results. At $L=256$, the fit is
morever hampered by the existence of oscillations that becomes more and more
important at large strip widths. At $L=128$ and $\beta=1.08$, the helicity
modulus can be extrapolated to the value $0.86450$ (with both power-law
and expoential fits) in the limit $m\rightarrow +\infty$. In the following,
$m=325$ states will be kept at each truncation. Since the helicity modulus takes
the value $0.86453$ for $m=325$, the systematic deviation at $L=128$ can be
estimated to be of order ${\cal O}(10^{-5})$. By the same procedure, the
deviation is shown to be one order of magnitude smaller for $L=64$.
For $L=256$, this systematic deviation is of order ${\cal O}(10^{-4})$.
In the following, numerical data for the 6-state clock will also be presented
for comparison. For this model, $m=648$ states were kept during the truncations
but strip widths only up to $L=64$ could be considered.

\section{Numerical evidence of two BKT transitions}
\subsection{Helicity modulus}
\begin{figure}
\centering
\psfrag{Y01}[Bc][Bc][1][1]{$\Upsilon_2$}
\psfrag{Y01}[Bc][Bc][1][1]{$\Upsilon_2$}
\psfrag{T}[tc][tc][1][0]{$T$}
\psfrag{L=16}[Bc][Bc][1][0]{\tiny $L=16$}
\psfrag{L=24}[Bc][Bc][1][0]{\tiny $L=24$}
\psfrag{L=32}[Bc][Bc][1][0]{\tiny $L=32$}
\psfrag{L=64}[Bc][Bc][1][0]{\tiny $L=64$}
\psfrag{L=128}[Bc][Bc][1][0]{\tiny $L=128$}
\psfrag{L=256}[Bc][Bc][1][0]{\tiny $L=256$}
\psfrag{L=512}[Bc][Bc][1][0]{\tiny $L=512$}
\includegraphics[width=\FigSize]{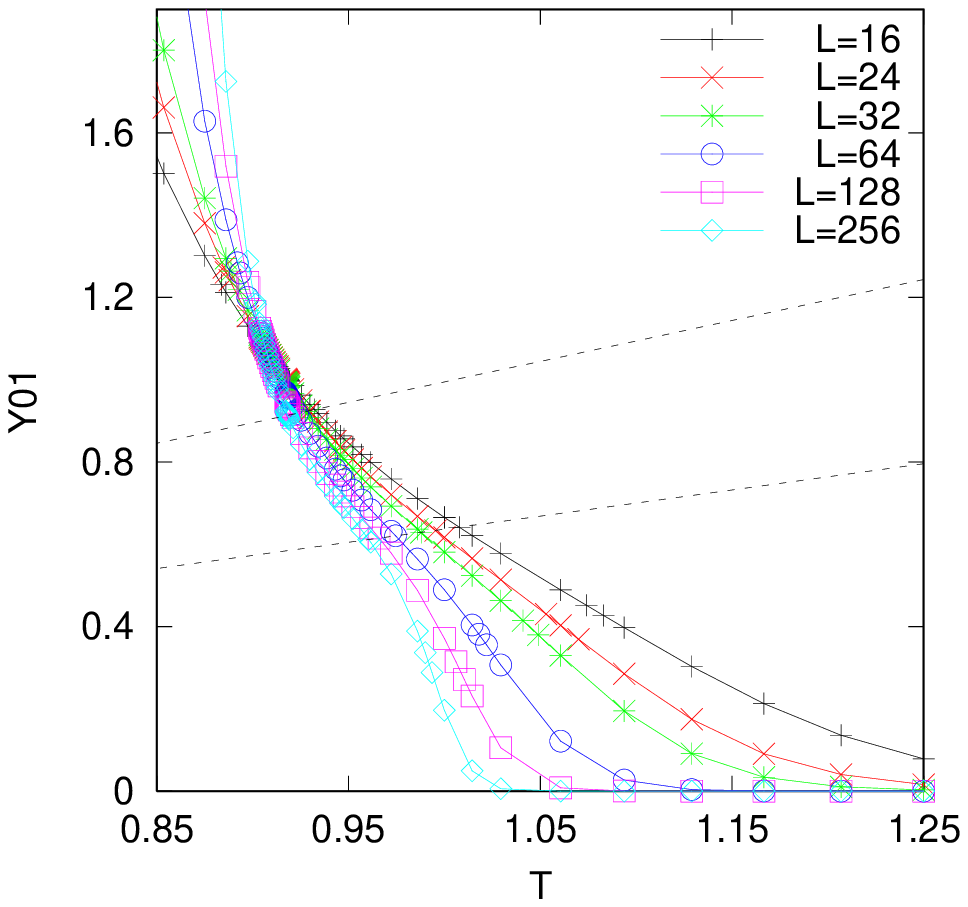}
\includegraphics[width=\FigSize]{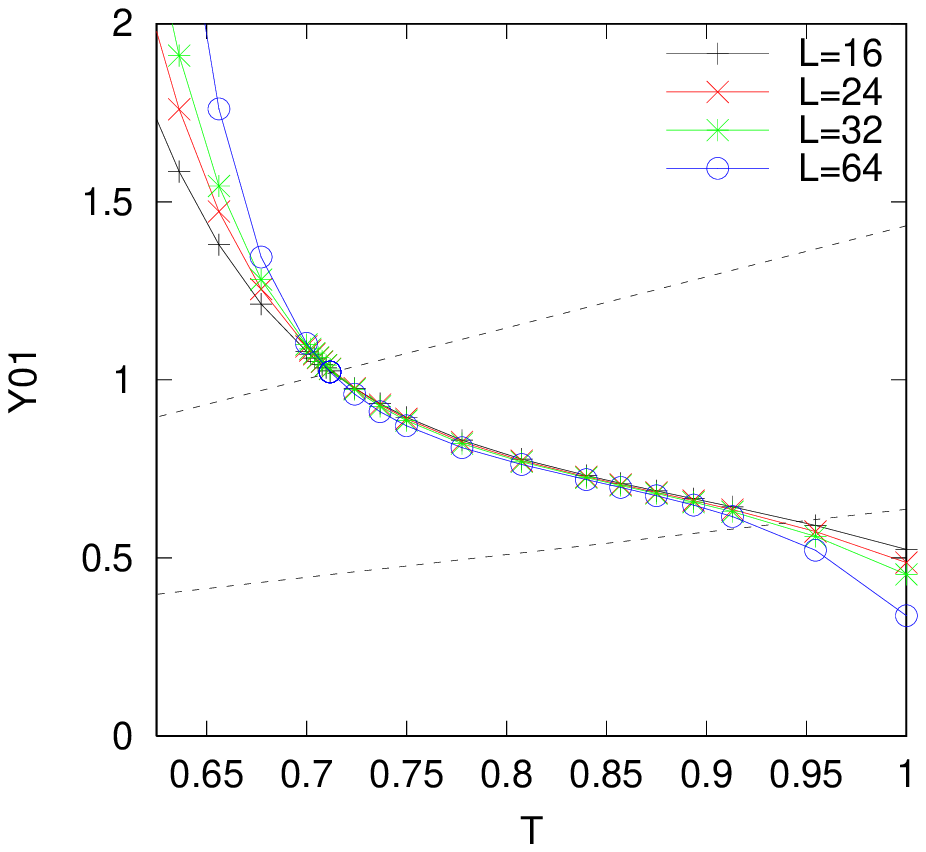}
\caption{On the left, helicity modulus $\Upsilon_2$ of the
5-state clock model versus the temperature $T$ for different
strip widths $L$. The two dashed lines correspond to the
predicted RG relations at the BKT transition temperatures.
On the right, the helicity modulus of the 6-state clock model
is plotted for comparison.}
\label{Fig2}
\end{figure}

A clear signature of the BKT transitions is provided by the
helicity modulus $\Upsilon_2$. The temperature dependence of $\Upsilon_2$
is plotted on figure~\ref{Fig2} for both the 5 and 6-state clock models.
In the ferromagnetic phase, the twist imposed at the boundaries induces
domain walls rather than spin waves. As a consequence, $L\big[f(\Delta)
-f(0)\big]$ is finite. Since the helicity modulus is proportional to
$L^2\big[f(\Delta)-f(0)\big]$, it diverges in the ferromagnetic phase,
as can be observed on the figure. In the paramagnetic phase, the correlation
length is finite so no (quasi) long-range order can propagate into the
system and the difference $L\big[f(\Delta)-f(0)\big]$, as well as the
helicity modulus, goes to zero as the strip width is increased. For
large strip widths, the helicity modulus is perfectly equal to zero,
in contradistinction to previous claims.
In the critical phase, the helicity modulus is expected to take a
size-independent value. If this behavior is clearly observed on
figure~\ref{Fig2} for the 6-state clock model, this is not yet the case
for the 5-state model, even though the curves come closer to each other
as the strip width is increased. The two expected jumps of helicity
modulus are not observed yet but the tangent of the curve becomes more
and more steep at temperatures close to the expected BKT transition
temperatures $T_{\rm BKT}^{\rm low}\simeq 0.90$ and $T_{\rm BKT}^{\rm high}
\simeq 0.95$.

\begin{figure}
\centering
\psfrag{T}[Bc][Bc][1][1]{$T$}
\psfrag{x}[Bc][Bc][1][1]{$1/(\ln L)^2$}
\psfrag{L=16}[Bc][Bc][1][0]{\tiny $L=16$}
\psfrag{L=24}[Bc][Bc][1][0]{\tiny $L=24$}
\psfrag{L=32}[Bc][Bc][1][0]{\tiny $L=32$}
\psfrag{L=64}[Bc][Bc][1][0]{\tiny $L=64$}
\psfrag{L=128}[Bc][Bc][1][0]{\tiny $L=128$}
\psfrag{L=256}[Bc][Bc][1][0]{\tiny $L=256$}
\psfrag{L=512}[Bc][Bc][1][0]{\tiny $L=512$}
\includegraphics[width=\FigSize]{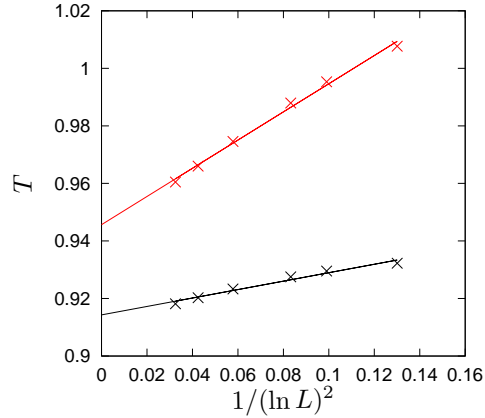}
\caption{Temperatures where the helicity modulus $\Upsilon_2$ of the
5-state clock model intersects the Renormalization Group predictions (\ref{UnivVal})
versus $1/(\ln L)^2$. $L$ is the width of the strip. The straight lines
are linear fits.}
\label{Fig2b}
\end{figure}

When the BKT transition point is approached from the non-critical
phase, the helicity modulus takes a universal value which is predicted
to be proportional to $T_{\rm BKT}$. In the case of the 2D clock model,
these two universal quantities are~\cite{Kumano}:
   \be \Upsilon_2(T_{\rm BKT}^{\rm high})={2\over\pi}T_{\rm BKT}^{\rm high},\quad\quad
     \Upsilon_2(T_{\rm BKT}^{\rm low})={q^2\over 8\pi}T_{\rm BKT}^{\rm low}.
     \label{UnivVal}\ee
These linear behaviors are plotted on figure~\ref{Fig2}. The different curves
are expected to intersect each one of these two straight lines at the
same point in the thermodynamic limit. As can be seen on the figure, and as already
noted by Kumano~{\sl et al.}~\cite{Kumano}, this is not yet the case in the
range of strip widths that were considered. However, the accuracy of the
DMRG data allows to determine the intersection of the helicity modulus
with the two predictions (\ref{UnivVal}). On figure~\ref{Fig2b},
the temperatures at which occur these intersections are plotted versus
$1/(\ln L)^2$. Since the correlation length displays an essential singularity
   \be \xi\sim e^{a|T-T_{\rm BKT}|^{-1/2}},\quad T>T_{\rm BKT}\label{SingXi}\ee
the temperature shift $T-T_{\rm KT}$ should indeed scale as $1/(\ln L)^2$
in the Finite-Size regime $\xi\sim L$. The data are in good agreement
with this statement. Linear fits lead then to the extrapolated BKT transition
temperatures $T_{\rm BKT}^{\rm low}\simeq 0.914(12)$ and $T_{\rm BKT}^{\rm high}
\simeq 0.945(17)$, in agreement with earlier estimates $T_{\rm BKT}^{\rm low}
\simeq 0.90514(9)$ and $T_{\rm BKT}^{\rm high}\simeq 0.95147(9)$~\cite{Papa}
obtained by other, more accurate, techniques but not involving the helicity
modulus.

\begin{figure}
\centering
\psfrag{T01}[Bc][Bc][1][1]{$(\beta-\beta_{\rm BKT}^{\rm high})\ \!(\ln L/L_0)^2$}
\psfrag{Y01}[Bc][Bc][1][1]{$(\pi\beta\Upsilon_2-2)\ln L/L_0$}
\psfrag{T02}[Bc][Bc][1][1]{$(\beta_{\rm BKT}^{\rm low}-\beta)\ \!(\ln L/L_0)^2$}
\psfrag{Y02}[Bc][Bc][1][1]{$(5^2-8\pi\beta\Upsilon_2)\ln L/L_0$}
\psfrag{L=16}[Bc][Bc][1][0]{\tiny $L=16$}
\psfrag{L=24}[Bc][Bc][1][0]{\tiny $L=24$}
\psfrag{L=32}[Bc][Bc][1][0]{\tiny $L=32$}
\psfrag{L=64}[Bc][Bc][1][0]{\tiny $L=64$}
\psfrag{L=128}[Bc][Bc][1][0]{\tiny $L=128$}
\psfrag{L=256}[Bc][Bc][1][0]{\tiny $L=256$}
\psfrag{L=512}[Bc][Bc][1][0]{\tiny $L=512$}
\includegraphics[width=\FigSize]{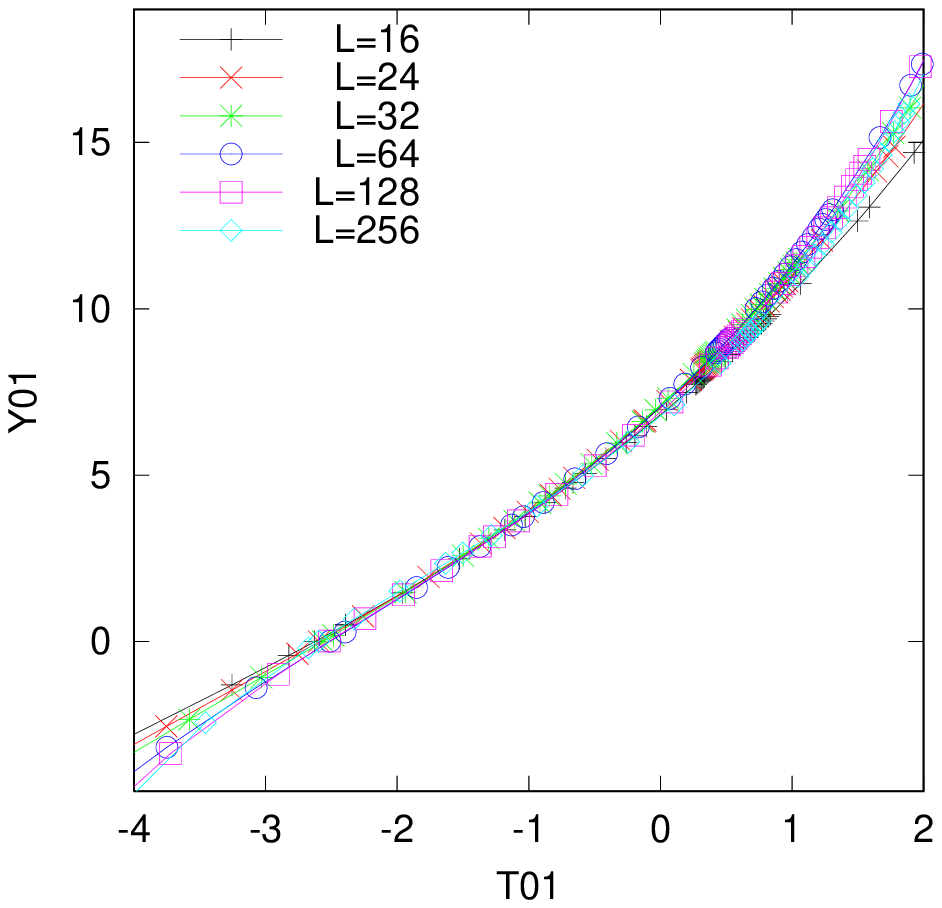}
\includegraphics[width=\FigSize]{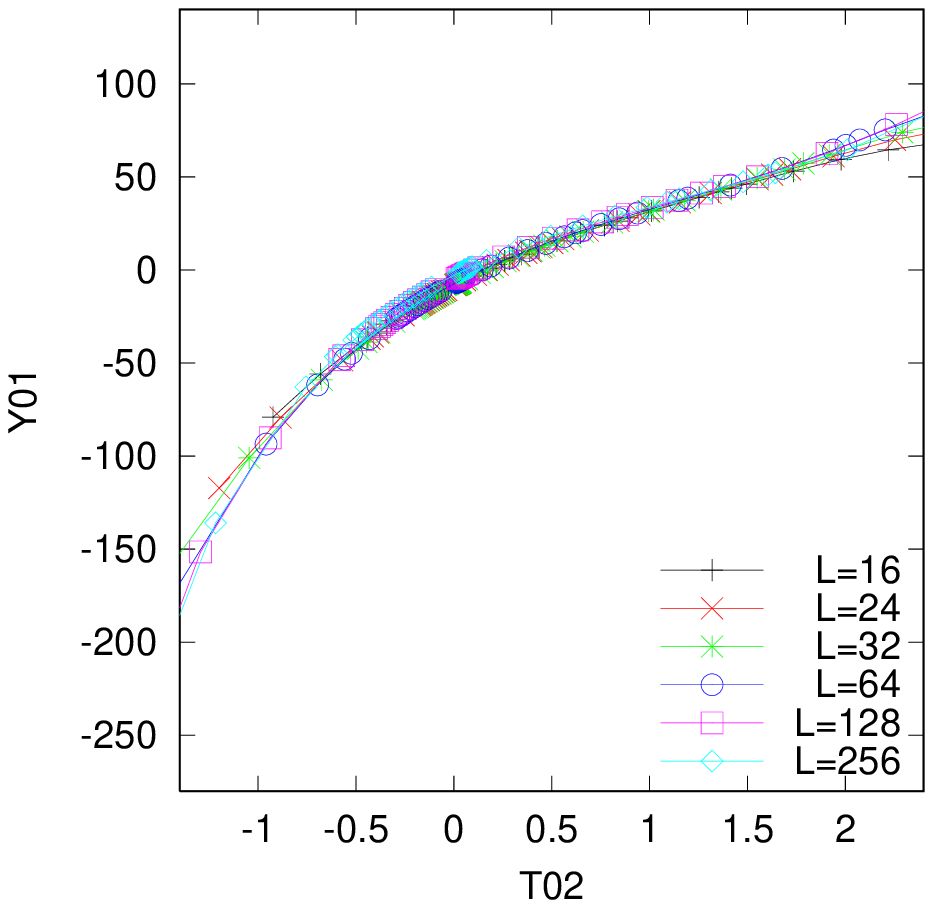}
\caption{Scaling function ${\cal F}_\Upsilon$ (up to a factor)
of the 5-state clock model with respect to the scaling argument
$(\beta_{\rm BKT}-\beta)\ \!(\ln L/L_0)^2$. The different
symbols correspond to different strip widths. The left
figure corresponds to the high-temperature BKT transition
while the right one corresponds to the low-temperature
transition.}
\label{Fig4}
\end{figure}

\begin{figure}
\centering
\psfrag{T01}[Bc][Bc][1][1]{$(\beta-\beta_{\rm BKT}^{\rm high})\ \!(\ln L/L_0)^2$}
\psfrag{Y01}[Bc][Bc][1][1]{$(\pi\beta\Upsilon_2-2)\ln L/L_0$}
\psfrag{T02}[Bc][Bc][1][1]{$(\beta_{\rm BKT}^{\rm low}-\beta)\ \!(\ln L/L_0)^2$}
\psfrag{Y02}[Bc][Bc][1][1]{$(5^2-8\pi\beta\Upsilon_2)\ln L/L_0$}
\psfrag{L=16}[Bc][Bc][1][0]{\tiny $L=16$}
\psfrag{L=24}[Bc][Bc][1][0]{\tiny $L=24$}
\psfrag{L=32}[Bc][Bc][1][0]{\tiny $L=32$}
\psfrag{L=64}[Bc][Bc][1][0]{\tiny $L=64$}
\psfrag{L=128}[Bc][Bc][1][0]{\tiny $L=128$}
\psfrag{L=256}[Bc][Bc][1][0]{\tiny $L=256$}
\psfrag{L=512}[Bc][Bc][1][0]{\tiny $L=512$}
\includegraphics[width=\FigSize]{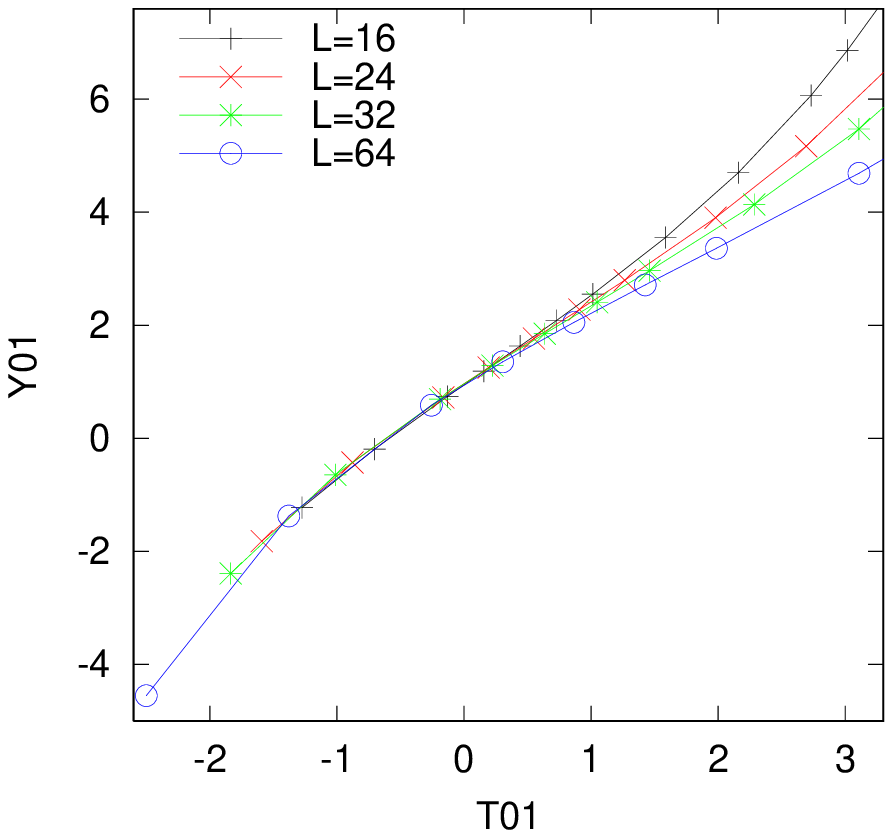}
\includegraphics[width=\FigSize]{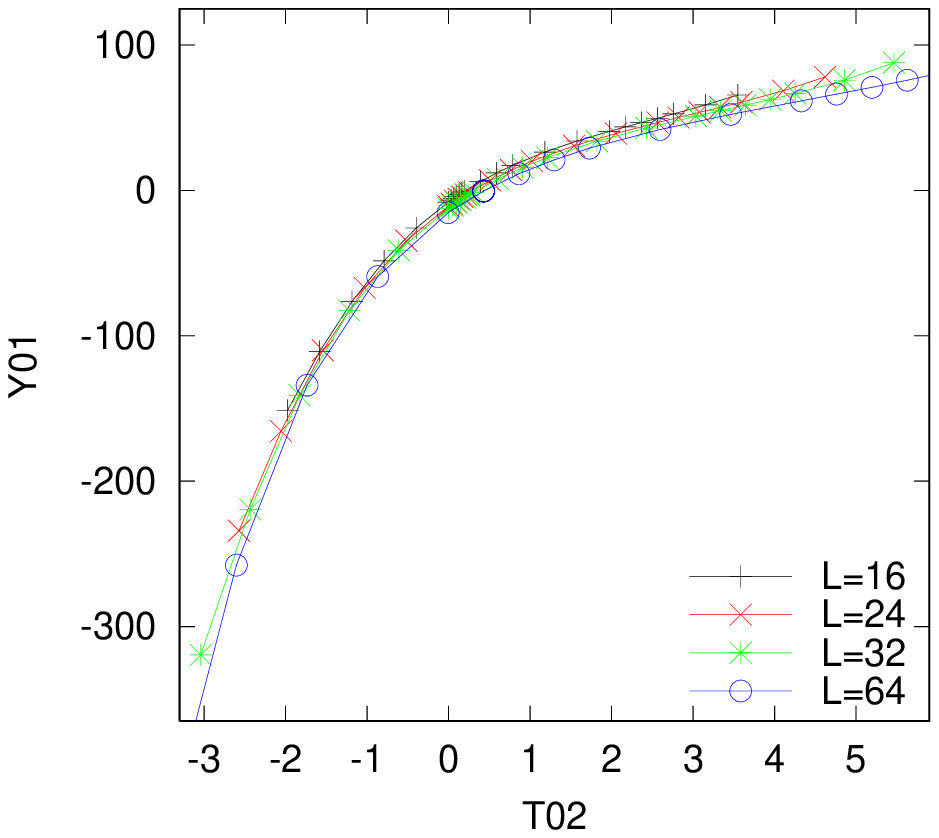}
\caption{Scaling function ${\cal F}_\Upsilon$ (up to a factor)
of the 6-state clock model with respect to the scaling argument
$(\beta_{\rm BKT}-\beta)\ \!(\ln L/L_0)^2$. The different
symbols correspond to different strip widths. The left
figure corresponds to the high-temperature BKT transition
while the right one corresponds to the low-temperature
transition.}
\label{Fig4b}
\end{figure}

A stronger prediction of Kosterlitz-Thouless theory is that
the universal value of the helicity modulus is approached as
    \be\Upsilon_2(T)-\Upsilon_2(T_{\rm BKT})\sim
    \Big[\ln{L\over L_0}\Big]^{-1}
    {\cal F}_\Upsilon\Big(|T-T_{\rm BKT}|(\ln L/L_0)^2\Big)\ee
where ${\cal F}_\Upsilon$ is a universal scaling function. 
This prediction is tested on figure~\ref{Fig4} for the two BKT
transitions of the clock model. The collapse of the curves is
reasonably good for a wide range of the scaling parameter.
For comparison, the same curves are plotted for the 6-state clock
model on figure~\ref{Fig4b}. In the case $q=5$, the transition
temperatures $T_{\rm BKT}^{\rm low}$ and $T_{\rm BKT}^{\rm high}$ were
treated, like $L_0$, as free parameters. The best values were defined
as the minimum of the square deviation around the mean over the
different strip widths. The collapse on figure~\ref{Fig4} was obtained
with $T_{\rm BKT}^{\rm low}\simeq 0.917$ and $L_0\simeq 0.87$ at the low
temperature BKT transition and $T_{\rm BKT}^{\rm high}\simeq 0.927$ and
$L_0\simeq 0.07$ at the second transition. These values are still relatively
different with the estimates of $T_{\rm BKT}^{\rm low}$ and $T_{\rm BKT}^{\rm low}$.
However, these values are very sensible to the interval of temperatures
on which the mean-square deviation is minimized. Almost as good collapses
can be obtained with the temperatures $T_{\rm BKT}^{\rm low}\simeq 0.90514(9)$
and $T_{\rm BKT}^{\rm high}\simeq 0.95147(9)$~\cite{Papa}. In the case
$q=6$ plotted on figure~\ref{Fig4b}, we restricted ourselves to
use the same values as in Ref.~\cite{Kumano}.

\subsection{Fourth-order helicity modulus}
\begin{figure}
\centering
\psfrag{ChiY}[Bc][Bc][1][1]{$\Upsilon_4$}
\psfrag{T}[tc][tc][1][0]{$T$}
\psfrag{x}[tc][tc][1][0]{$1/(\ln L)^2$}
\psfrag{L=16}[Bc][Bc][1][0]{\tiny $L=16$}
\psfrag{L=24}[Bc][Bc][1][0]{\tiny $L=24$}
\psfrag{L=32}[Bc][Bc][1][0]{\tiny $L=32$}
\psfrag{L=64}[Bc][Bc][1][0]{\tiny $L=64$}
\psfrag{L=128}[Bc][Bc][1][0]{\tiny $L=128$}
\psfrag{L=256}[Bc][Bc][1][0]{\tiny $L=256$}
\psfrag{L=512}[Bc][Bc][1][0]{\tiny $L=512$}
\includegraphics[width=\FigSize]{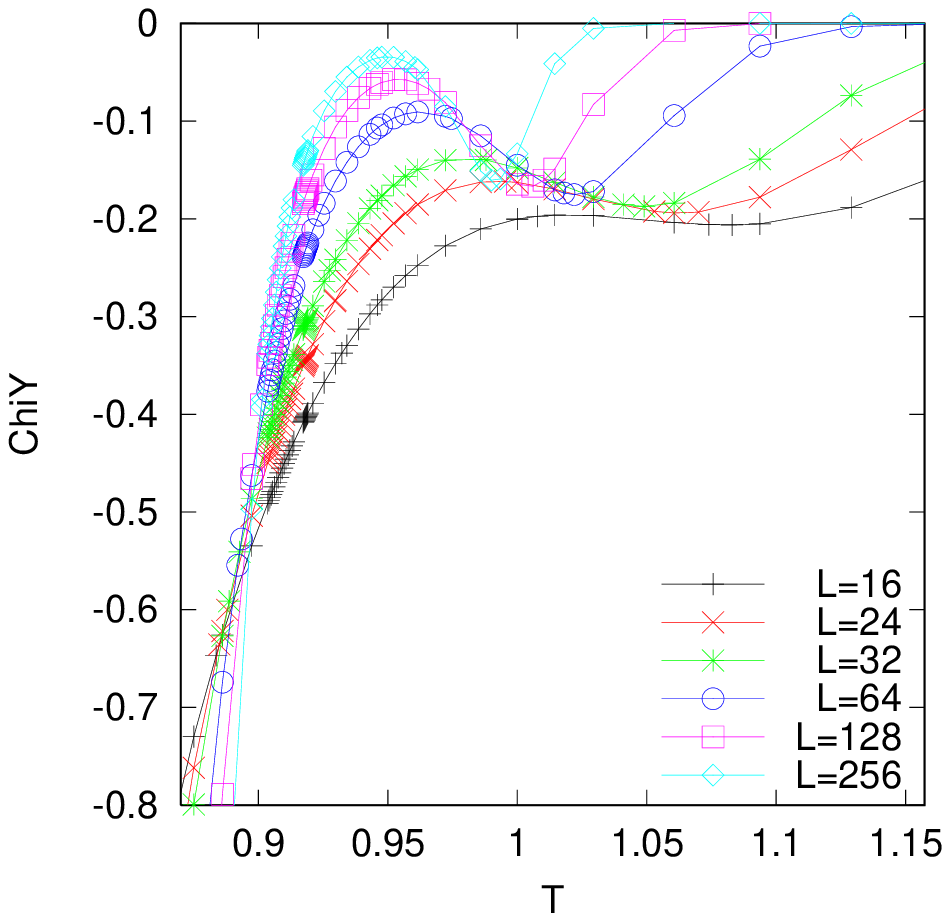}
\includegraphics[width=\FigSize]{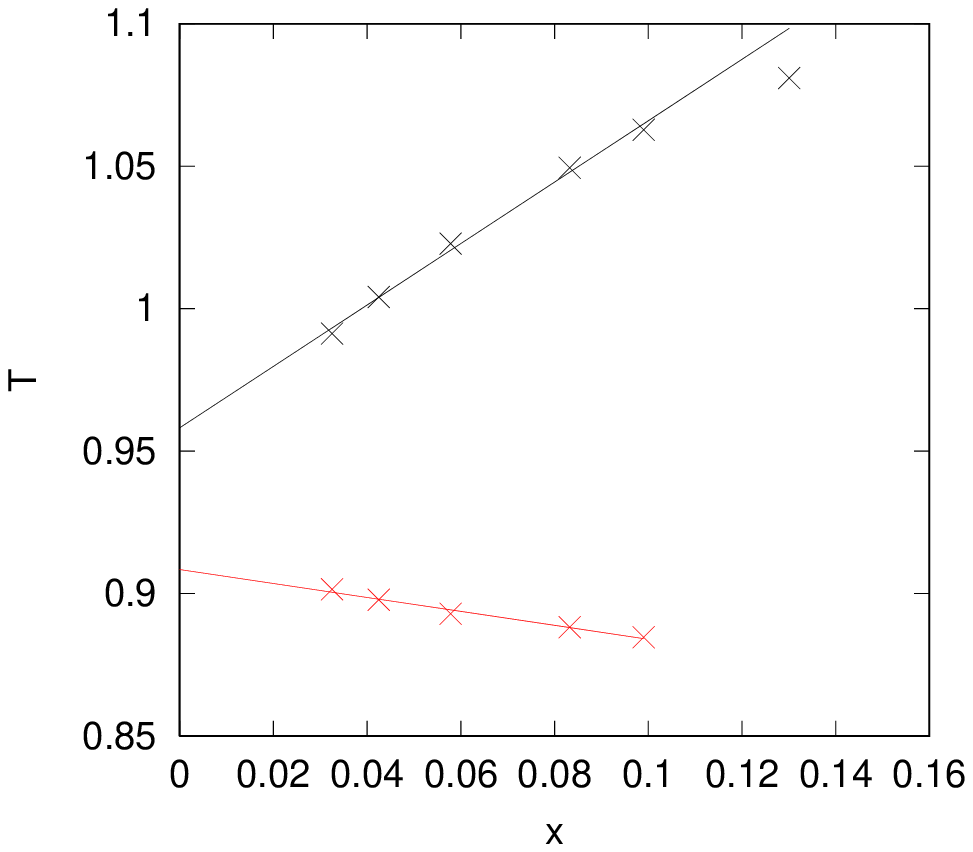}
\caption{On the right, fourth derivative $\Upsilon_4$ of the free energy
density (Eq. \ref{Y4}) of the 5-state clock model versus temperature $T$.
The different curves correspond to different strip widths $L$.
The two BKT transitions are signaled by respectively the crossing of
the curves and the dip. On the left, temperatures at which the intersection
of the curves are observed (red) and location of the dip (black) versus
$1/(\ln L)^2$. The straight lines are linear fits.}
\label{Fig3}
\end{figure}

As shown by Kim {\sl et al.} in the case of the XY model~\cite{Kim03},
the discontinuity of the helicity modulus $\Upsilon_2$ at the BKT transition
temperature manifests itself as a dip in higher-order derivatives of the
free energy density. The temperature
dependence of the fourth-order helicity modulus $\Upsilon_4$, estimated as
(\ref{Y4}), is plotted on figure~\ref{Fig3}. Like the helicity modulus
$\Upsilon_2$, it vanishes in the high temperature phase. The formation of a
dip is clearly seen around $T_{\rm BKT}^{\rm high}$.
However, the low-temperature BKT transition does not
manifest itself as a dip. Nevertheless, it seems that the intersection
of the curves at temperatures close to $T_{\rm BKT}^{\rm low}$ indicates that
$\Upsilon_4$ takes a universal value, like $\Upsilon_2$, at $T_{\rm BKT}^{\rm low}$
rather than displaying a dip. On the right figure, the location of the dip
and the temperatures at which the curves intersect 
for two successive strip widths are plotted versus $1/(\ln  L)^2$.
Both can only be roughly determined but, nevertheless, a linear behavior
is observed if one excludes the point $L=16$. Linear fits give the
BKT transition temperatures $T_{\rm BKT}^{\rm low}\simeq 0.91(2)$ and
$T_{\rm BKT}^{\rm high}\simeq 0.96(4)$, again in agreement with earlier estimates
$T_{\rm BKT}^{\rm low}\simeq 0.90514(9)$ and $T_{\rm BKT}^{\rm high}\simeq
0.95147(9)$~\cite{Papa}.

\subsection{Magnetization}
\begin{figure}
\centering
\psfrag{logM2}[Bc][Bc][1][1]{$(\ln M)^2$}
\psfrag{M}[Bc][Bc][1][1]{$m$}
\psfrag{T}[tc][tc][1][0]{$T$}
\psfrag{x}[Bc][Bc][1][1]{$(\beta_{\rm BKT}^{\rm low}-\beta)\ \!(\ln L/L_0)^2$}
\psfrag{y}[tc][tc][1][0]{$m\ \!L^{1/8}$}
\psfrag{L=16}[Bc][Bc][1][0]{\tiny $L=16$}
\psfrag{L=24}[Bc][Bc][1][0]{\tiny $L=24$}
\psfrag{L=32}[Bc][Bc][1][0]{\tiny $L=32$}
\psfrag{L=64}[Bc][Bc][1][0]{\tiny $L=64$}
\psfrag{L=128}[Bc][Bc][1][0]{\tiny $L=128$}
\psfrag{L=256}[Bc][Bc][1][0]{\tiny $L=256$}
\psfrag{L=512}[Bc][Bc][1][0]{\tiny $L=512$}
\includegraphics[width=\FigSize]{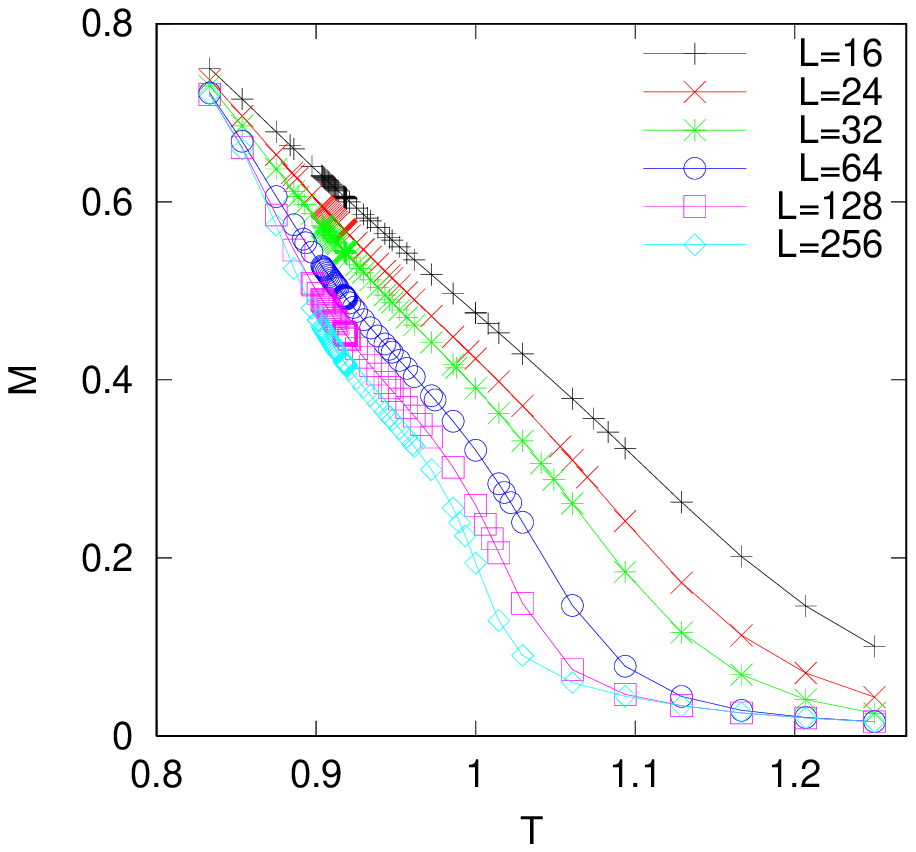}
\includegraphics[width=\FigSize]{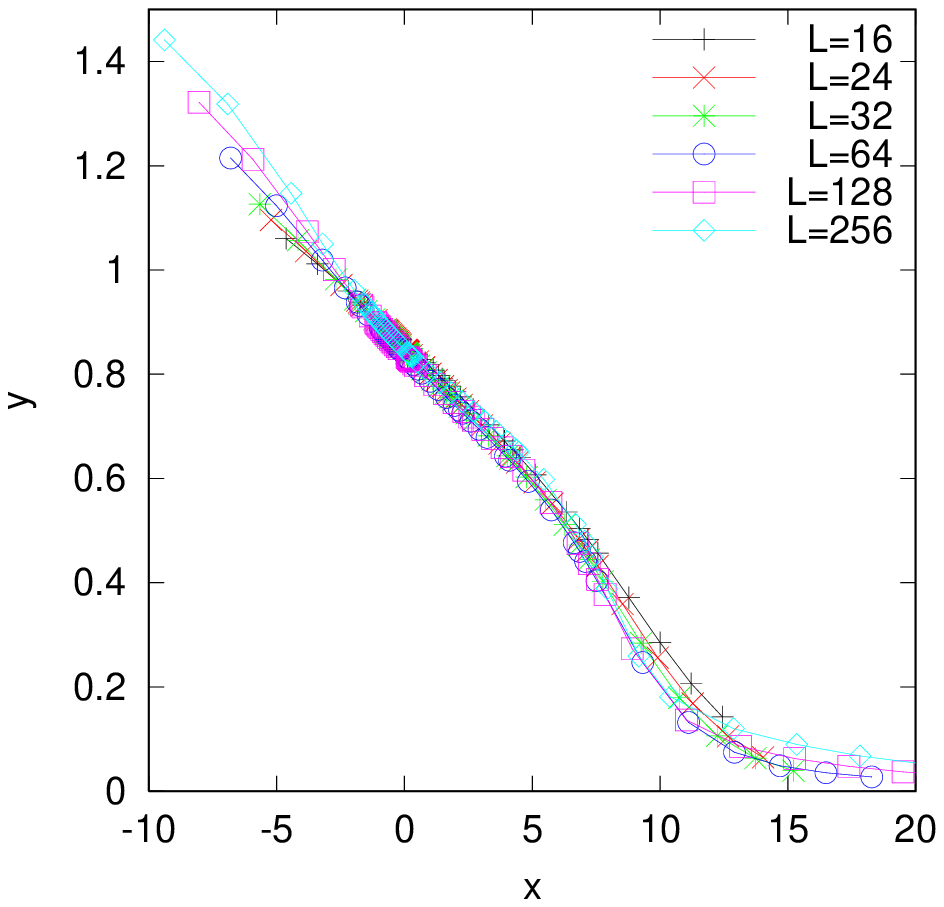}
\caption{On the left, magnetization density $m$ of the
5-state clock model versus the temperature $T$ for different
strip widths $L$. On the right, the scaling function
$m\ L^{1/8}$ is plotted with respect to the scaling parameter
$(\beta_{\rm BKT}^{\rm low}-\beta)\ \!(\ln L/L_0)^2$.
}\label{Fig1}
\end{figure}

Even though the system is infinite in the longitudinal direction of the
strip, a finite spontaneous magnetization is observed because magnetic
fields are applied at the boundaries. The magnetization $m$ is measured
on the site $L/2$, i.e. at the center of the strip. The data are plotted on
figure~\ref{Fig1}. In the scaling regime $L\gg\xi$, $m$ is expected to
display an essential singularity $\xi^{-\eta/2}$ while, in the finite-size
regime, the usual algebraic law with the lattice size, $m\sim L^{-\beta/\nu}$
with $\beta/\nu=\eta/2$, is recovered. The magnetic scaling dimension is
usually conjectured to take the same value as in the XY model, i.e. $\eta=1/4$,
even though recent Monte Carlo simulations systematically led to smaller
values~\cite{Papa}. On figure~\ref{Fig1}, the scaling hypothesis
    \be m\sim L^{-\eta/2}{\cal F}_m(|T-T_{\rm BKT}|\ \!(\ln L/L_0)^2)
    \label{ScalMag}\ee
is tested with the value $\eta/2=1/8$. The collapse of the curves corresponding
to different strip widths on figure~\ref{Fig1} (right) shows that the scaling
function ${\cal F}_m$ depends, as expected, only on $|T-T_{\rm BKT}|\ \!(\ln L/L_0)^2$.
Again, the mean-square deviation was minimized to find to optimal values
of the free parameters $T_{\rm BKT}^{\rm low}$ and $L_0$. The values
$T_{\rm BKT}^{\rm low}\simeq 0.917$ and $L_0\simeq 0.023$ were obtained.
As in the case of the helicity modulus, small joint variations of these
parameters lead to almost as good collapses so the method is not reliable
to determine $T_{\rm BKT}^{\rm low}$ accurately. However, it provides evidence
of the scaling behavior~(\ref{ScalMag}). We also tried to extract the magnetic
exponent $\eta/2$ from a fit of magnetization as $m\sim L^{-\eta/2}$. At the
temperature $T_{\rm BKT}^{\rm low}\simeq 0.90514$, the estimated exponent is
$0.116(4)$, i.e. slightly below the expected value $1/8$. However, it increases
with temperature, which suggests that strong scaling corrections are present.
Using the temperature $T_{\rm BKT}^{\rm low}\simeq 0.917$ that leads to the best scaling
Eq.~\ref{ScalMag}, the exponent takes the value $\eta/2=0.1254(30)$, in
perfect agreement with the expected value.

\section{Conclusions}
The two BKT phase transitions of the 5-state clock model were studied
using the DMRG algorithm. The latter is shown to allow for larger
lattice sizes with a better accuracy than Monte Carlo simulations.
Despite the Open Boundary Conditions imposed by the technique, 
the helicity modulus $\Upsilon_2$, as well as higher-derivative
$\Upsilon_4$, can be computed by imposing twisted magnetic fields
on the two boundaries of the strip. The topological nature of the
two phase transitions is confirmed. Finite-Size displacements of
the BKT transition temperatures scale as $1/(\ln L)^2$,
in agreement with an essential singularity of the correlation
length. The exponent $\sigma$ is therefore the same as for the
XY model for both BKT transitions. The helicity modulus is shown
to tend towards the universal values predicted by Renormalization Group
techniques at the temperatures compatible with the most accurate
Monte Carlo estimates. Finally, the scaling behavior of magnetization
is also compatible with an essential singularity with the same magnetic
scaling dimension $\eta/2=1/8$ as the XY model.

\section*{References}

\end{document}